\documentclass[twoside,english]{iopart}
\usepackage[T1]{fontenc}
\usepackage[latin9]{inputenc}
\usepackage{geometry}
\usepackage{array}
\usepackage{booktabs}
\usepackage{graphicx}

\makeatletter

\providecommand{\tabularnewline}{\\}

\usepackage{iopams}
\usepackage{setstack}

\usepackage[numbers, sort&compress]{natbib}

\newcommand{\eqref}[1]{(\ref{#1})}

\usepackage[singlelinecheck=on,justification=justified]{caption}
\usepackage[load-configurations=abbreviations,range-phrase = ~-~,range-units = single,separate-uncertainty = true]{siunitx}
\usepackage[siunitx]{circuitikz}
\RequirePackage{subscript}
\usepackage{ifpdf}
\ifpdf
\usepackage{epstopdf}
\fi
\usepackage[labelfont=bf]{caption}
\newcommand{\YBCO}{\emph{RE}BCO~}

\newcommand{\sym}[2]{\ensuremath{#1_\textrm{#2}}}

\newcommand{\n}[1]{\textrm{#1}}

\makeatother

\usepackage{babel}
\begin{document}

\title[Temperature- and Field Dependent Characterization of a Twisted Stacked-Tape
Cable]{Temperature- and Field Dependent Characterization of a Twisted Stacked-Tape
Cable.}

\author{{\Large{}C Barth{\Large{}\textsuperscript{1,2}}, M Takayasu{\Large{}\textsuperscript{3}},
N Bagrets{\Large{}\textsuperscript{1}}, C M Bayer{\Large{}\textsuperscript{1}},
K-P Weiss{\Large{}\textsuperscript{1}} and C Lange{\Large{}\textsuperscript{1}} }}

\address{{\large{}{\large{}\textsuperscript{1}} Institute for Technical Physics
(ITEP), Karlsruhe Institute of Technology (KIT), Germany}}

\address{{\large{}{\large{}\textsuperscript{2}} Department of Condensed Matter
Physics (DPMC), University of Geneva, Switzerland}}

\address{{\large{}{\large{}\textsuperscript{3}} Plasma Science and Fusion
Center (PSFC), Massachusetts Institute of Technology (MIT), USA}}

\ead{{\large{}christian.barth@unige.ch}}
\begin{abstract}
The Twisted Stacked-Tape Cable (TSTC) is one of the major high temperature
superconductor cable concepts combining scalability, ease of fabrication
and high current density making it a possible candidate as conductor
for large scale magnets. To simulate the boundary conditions of such
a magnets as well as the temperature dependence of Twisted Stacked-Tape
Cables a \SI{1.16}{\metre} long sample consisting of 40, \SI{4}{\milli\metre}
wide SuperPower \YBCO tapes is characterized using the ``FBI''
(force - field - current) superconductor test facility of the Institute
for Technical Physics (ITEP) of the Karlsruhe Institute of Technology
(KIT). In a first step, the magnetic background field is cycled while
measuring the current carrying capabilities to determine the impact
of Lorentz forces on the TSTC sample performance. In the first field
cycle, the critical current of the TSTC sample is tested up to \SI{12}{\tesla}.
A significant Lorentz force of up to \SI{65.6}{\kilo\newton\per\metre}
at the maximal magnetic background field of \SI{12}{\tesla} result
in a \SI{11.8}{\percent} irreversible degradation of the current
carrying capabilities. The degradation saturates (critical cable current
of \SI{5.46}{\kilo\ampere} at \SI{4.2}{\kelvin} and \SI{12}{\tesla}
background field) and does not increase in following field cycles.
In a second step, the sample is characterized at different background
fields (\SIrange{4}{12}{\tesla}) and surface temperatures (\SIrange{4.2}{37.8}{\kelvin})
utilizing the variable temperature insert of the ``FBI'' test facility.
In a third step, the performance along the length of the sample is
determined at \SI{77}{\kelvin}, self-field. A \SI{15}{\percent}
degradation is obtained for the central part of the sample which was
within the high field region of the magnet during the in-field measurements.
\end{abstract}


\submitto{\SUST }

\maketitle

\section{Introduction}

Second generation high temperature superconductors (HTS) are the rare-earth-barium-copper-oxide
(\emph{RE}BCO) tapes, often referred to as coated conductors. They
are of thin tape shape, commonly with widths of \SIrange{3}{15}{\milli\metre}
and thicknesses in the \SIrange{50}{200}{\micro\metre} range. The
mechanical properties as well as the performance in strong magnetic
background fields surpasses first generation high temperature superconductors
making \YBCO tapes a desired conductor for rotating machinery, fusion
magnets and high field magnets.

\subsection{High temperature superconductor cable concepts}

Due to their tape shape, the established cabling methods employed
for low temperature superconductors are not applicable. Different
approaches are necessary. Around the world, there are at present five
major high temperature superconductor cable concepts of how to combine
several \YBCO tapes into flexible, mechanically strong cables able
to carry \si{\kilo\ampere} currents in strong background fields. 
\begin{description}
\item [{Roebel~Assembled~Coated~Conductor~(RACC)~cables}] are being
developed at the Karlsruhe Institute of Technology (KIT), Germany
and the Robinson Research Institute of the Victoria University of
Wellington, New Zealand \cite{Goldacker2006,Long2010,Long2011,Long2012}. 
\item [{Coated~Conductor~Rutherford~Cables~(CCRC)}] are being developed
at the Karlsruhe Institute of Technology (KIT), Germany \cite{Schlachter2011,Kario2013}.
\item [{Conductor~on~Round~Core~(CORC)~cables}] are being developed
at Advanced Conductor Technologies, USA, and the University of Colorado,
USA \cite{VanderLaan2009,VanderLaan2011a,VanderLaan2012a,VanderLaan2013}. 
\item [{Twisted~Stacked-Tape~Cables~(TSTC)}] are being developed at
the Massachusetts Institute of Technology, USA \cite{Takayasu2009,takayasu2011,Takayasu2012,Takayasu2012a,Chiesa2014}
and at the Italian National Agency for New Technologies, Energy and
Sustainable Economic Development (ENEA), Italy \cite{Celentano2014}.
\item [{Round~Strands~Composed~of~Coated~Conductor~Tapes~(RSCCCT)}] are
being developed at the Centre de Recherches en Physique des Plasmas
(CRPP) of the \'Ecole Polytechnique F\'ed\'erale de Lausanne (EPFL)
\cite{Uglietti2013}. 
\end{description}
There are significant differences in the arrangement of the tapes,
the tape consumption, the transposition, the mechanical properties
as well as the in-field performance of the HTS cable concepts. Their
applicability is thus depending on each application's boundary conditions.
An overview of HTS cable concepts can be found in \cite{Barth-PhD}.

\subsection{Twisted Stacked-Tape Cables\label{sub:TSTC}}

The Twisted Stacked-Tape Cable (TSTC) is a high temperature superconductor
(HTS) cable concept proposed by M. Takayasu et al. in 2009 \cite{Takayasu2009,takayasu2011}
to provide ``simple, high current density and scalable cabling method
applicable to a large scale magnet'' \cite{Takayasu2012}.

\begin{figure}[tbh]
\begin{centering}
\includegraphics[width=0.8\textwidth]{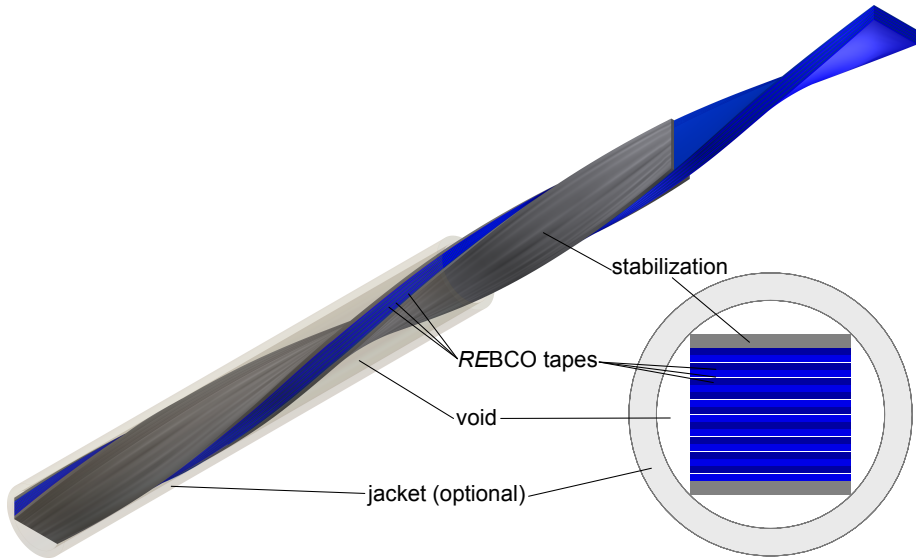}
\par\end{centering}

\caption{Schematic drawing of a Twisted Stacked-Tape Cable (TSTC). \YBCO tapes
are stacked and twisted. The mechanical and electrical stabilization
is increased with copper tapes on the top and the bottom of the superconductor
stack. To further improve the mechanical properties, the TSTC can
be equipped with a jacket of structural material.\label{fig:TSTC}}
\end{figure}

In TSTCs, several rare-earth-barium-copper-oxide (\emph{RE}BCO) tapes
are stacked and twisted. \SI{4}{\milli\metre} wide tapes are commonly
used; these are mechanically stabilized with copper tapes of the same
width, which are attached to the top and the bottom of the stack.
For additional mechanical stabilization, the twisted stack can be
inserted into a tube (jacket) of structural material forming a cable-in-conduit
conductor (CICC). To prevent movement of the tapes and to avoid stress
concentrations, all the voids between the TSTC and the sheath have
to be filled using glues, resins or solders. This HTS cable type is
shown schematically in figure~\ref{fig:TSTC}.

\section{Sample parameters\label{sec:Sample-parameters}}

A Twisted Stacked-Tape Cable sample of \SI{1.16}{\metre} length consisting
of 40, \SI{4}{\milli\metre} wide copper stabilized SuperPower \YBCO
tapes with advanced pinning (\SI{7.5}{at\percent} Zr doping) is used
in the following investigations. The tapes are arranged in one stack
to which \SI{0.25}{\milli\metre} thick copper tapes of the same width
are attached to the top and the bottom for mechanical and thermal
stabilization. The stack is twisted with a twist pitch of approximately
\SI{200}{\milli\metre} and is inserted into a copper tube of \SI{7.9}{\milli\metre}
outer diameter and \SI{0.81}{\milli\metre} wall thickness. The hollow
space between the stack and the tube is completely filled with soft
solder preventing movement of the stack and dispersing mechanical
loads. The sample contains two clamped BSCCO-\YBCO connectors at
its ends, which are described in detail in \cite{Takayasu2012a}.
Four pairs of voltage taps are soldered onto the copper tube and one
pair is attached to the copper terminations of the sample. All sample
parameters are summarized in table~\ref{tab:Sample-parameters}.

\begin{table}[tbh]
\caption{Parameters of the investigated Twisted Stacked-Tape Cable sample.\label{tab:Sample-parameters}}

\centering{}%
\begin{tabular}{cc}
\toprule 
{\footnotesize{}parameter} & {\footnotesize{}TSTC sample}\tabularnewline
\midrule
\midrule 
{\footnotesize{}sample length} & {\footnotesize{}\SI{1.16}{\metre} including terminations}\tabularnewline
\midrule 
{\footnotesize{}superconductor} & {\footnotesize{}\SI{4}{\milli\metre} wide copper stabilized from
SuperPower (SCS4050) - advanced pinning}\tabularnewline
\midrule 
{\footnotesize{}number of tapes} & {\footnotesize{}40}\tabularnewline
\midrule 
{\footnotesize{}twist pitch $\tau$} & {\footnotesize{}\SI{200}{\milli\metre}}\tabularnewline
\midrule 
{\footnotesize{}termination} & {\footnotesize{}clamped BSCCO - {\footnotesize{}\YBCO} connections}\tabularnewline
\midrule 
{\footnotesize{}mechanical stabilization} & {\footnotesize{}twisted stack soldered into Cu tube (\SI{7.90}{\milli\metre}
diameter, \SI{0.81}{\milli\metre} wall thickness)}\tabularnewline
\midrule 
{\footnotesize{}electrical stabilization} & {\footnotesize{}Cu stabilization of \YBCO tapes + Cu tapes on top
and bottom of the stack}\tabularnewline
\midrule 
{\footnotesize{}voltage taps} & {\footnotesize{}4 pairs soldered onto the Cu tube, 1 pair at the Cu
terminations }\tabularnewline
\bottomrule
\end{tabular}
\end{table}

To determine the impact of transverse (radial direction) mechanical
loads, magnetic background fields and temperature on the current carrying
capabilities of Twisted Stacked-Tape Cables, the TSTC sample is characterized
in three steps.
\begin{enumerate}
\item Impact of transverse mechanical loads by cycling the magnetic background
fields and measuring the current carrying capabilities. Degradations
due to the increasing Lorentz forces will be made visible in these
measurements by comparing the obtained critical currents of different
load cycles (subsection~\ref{sec:Load-cycling}).
\item Critical current measurements at different fields and temperatures
to determine the magnetic field and temperature dependence of Twisted
Stacked-Tape Cables (subsection~\ref{sec:T-variable}).
\item Critical current measurements at \SI{77}{\kelvin}, self-field after
the in-field characterizations to determine the irreversible influence
of the Lorentz forces onto the cable performance in different sections
along the length of the sample (subsection~\ref{sec:77K-characterization}).
\end{enumerate}

\section{Variable temperature insert\label{sec:Test-setup}}

\begin{figure}[tbh]
\begin{centering}
\includegraphics[width=1\textwidth]{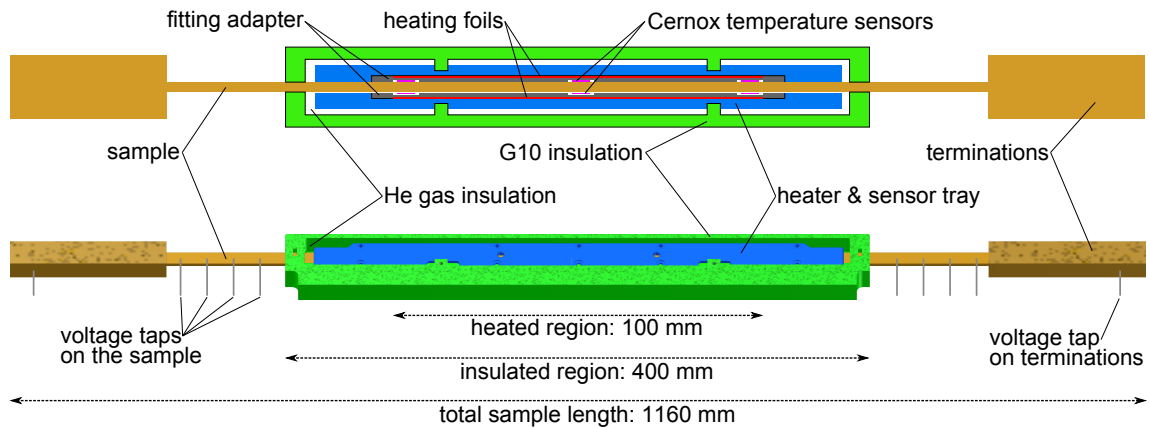}
\par\end{centering}

\caption{Variable temperature insert for the FBI test facility. The insert
allows testing at sample temperatures from \SIrange{4.2}{77}{\kelvin}.
Schematic drawing in side view (upper), technical drawing (lower).
Picture is not up to scale.\label{fig:FBI-t-variable-insert}}
\end{figure}

The experiments are performed using the FBI (force - field - current)
superconductor test facility of the Institute for Technical Physics
(ITEP) of the Karlsruhe Institute of Technology (KIT) \cite{Bayer2014}.
The name ``FBI'' of the test facility is an abbreviation of the
possible test parameters, \textbf{F} stands for force, \textbf{B}
for magnetic field and \textbf{I} for current. It contains a superconducting
split coil magnet, delivering magnetic background fields up to \SI{12}{\tesla},
the field is orientated perpendicular to the sample axis with a homogeneous
region (\SI{97}{\percent} of peak field) of \SI{70}{\milli\metre}.
A low noise direct current power supply delivers up to \SI{10}{\kilo\ampere}
to the sample. The test facility has been equipped with a variable
temperature insert \cite{Barth2011-MEM}, as shown schematically in
figure~\ref{fig:FBI-t-variable-insert}, to allow testing at sample
temperatures above \SI{4.2}{\kelvin}. It consists of:
\begin{description}
\item [{temperature~sensors:}] Two Cernox\texttrademark temperature sensors
\cite{Cernox-SUST} in direct contact with the sample to determine
its surface temperature. Cernox sensors are used as this sensor type
combines a high sensitivity at cryogenic temperatures with a very
low magnetic field dependence.
\item [{heating~foils:}] Two Kapton\textsuperscript{\textregistered}
heating foils of \SI{100}{\milli\metre} length each wrapped around
the sample. The heating power of the foils is controlled individually.
\item [{thermal~insulation:}] Double wall glass-fiber-reinforce-plastics
(GFRP or G10) chamber of \SI{400}{\milli\metre} length, containing
gaseous helium during operation. The chamber fits tightly around the
sample and is sealed at the upper and lower ends with adhesive tape
and bees wax in order to minimizes the helium exchange.
\end{description}

\begin{figure}[tbh]
\begin{centering}
\includegraphics[width=0.25\textwidth]{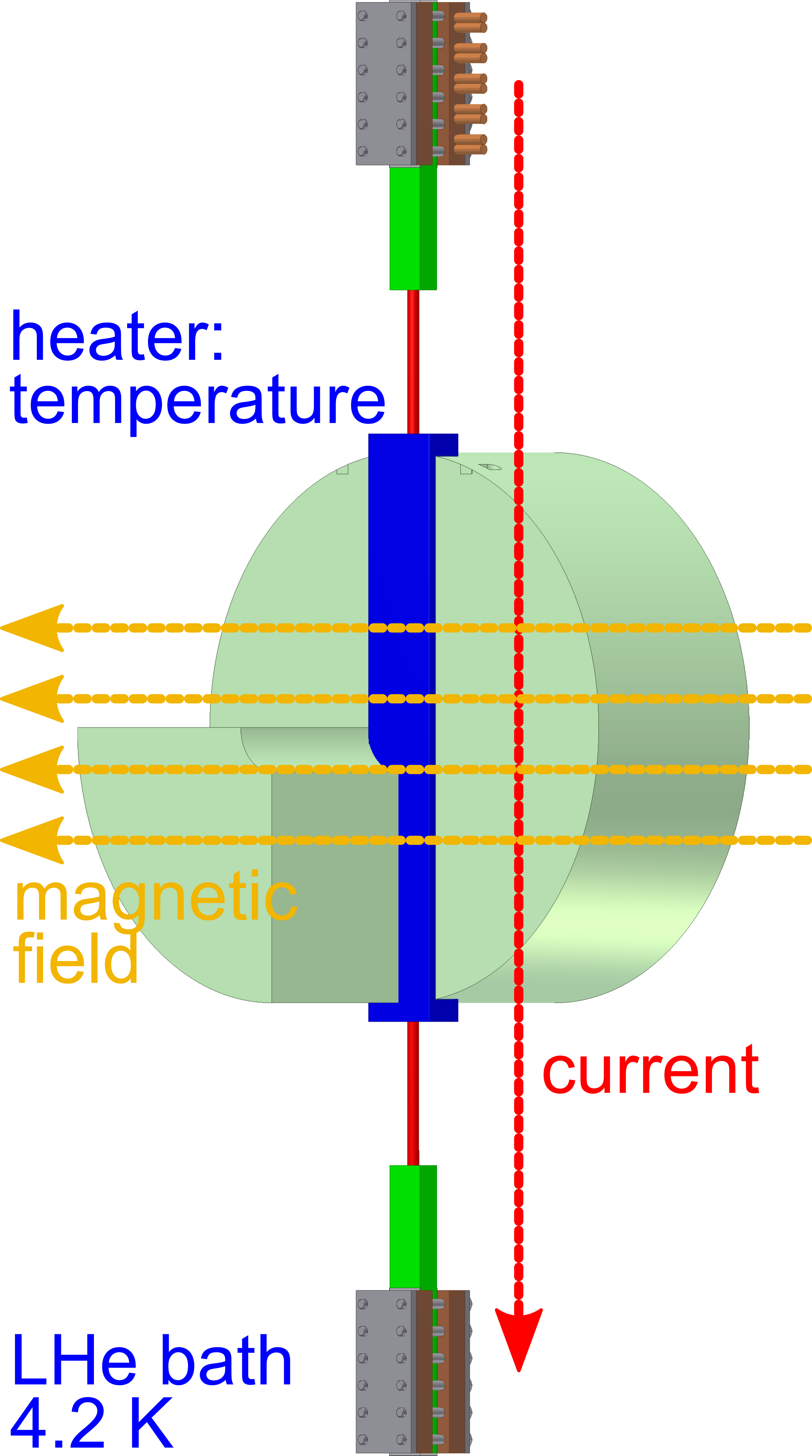}
\par\end{centering}

\caption{Schematic drawing of the superconductor test facility ``FBI'' with
the variable temperature variable. The system allows critical current
measurements on superconductor cables depending on magnetic background
fields and temperatures.\label{fig:Test-method}}
\end{figure}

The measurement configuration used for the characterization of the
Twisted Stacked-Tape Cable sample is shown schematically in figure~\ref{fig:Test-method}.

\subsection{Validation of the variable temperature insert\label{sub:T-variable-insert-validation}}

\begin{figure}[tbh]
\begin{centering}
\begin{tabular}{>{\centering}m{0.45\textwidth}>{\raggedleft}m{0.5\textwidth}}
\includegraphics[width=0.33\textwidth]{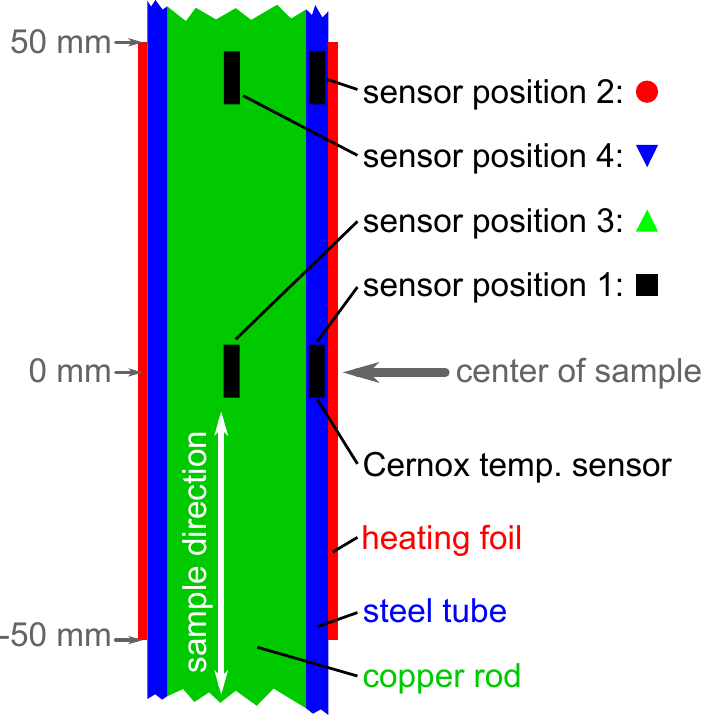} & \includegraphics[width=0.5\textwidth]{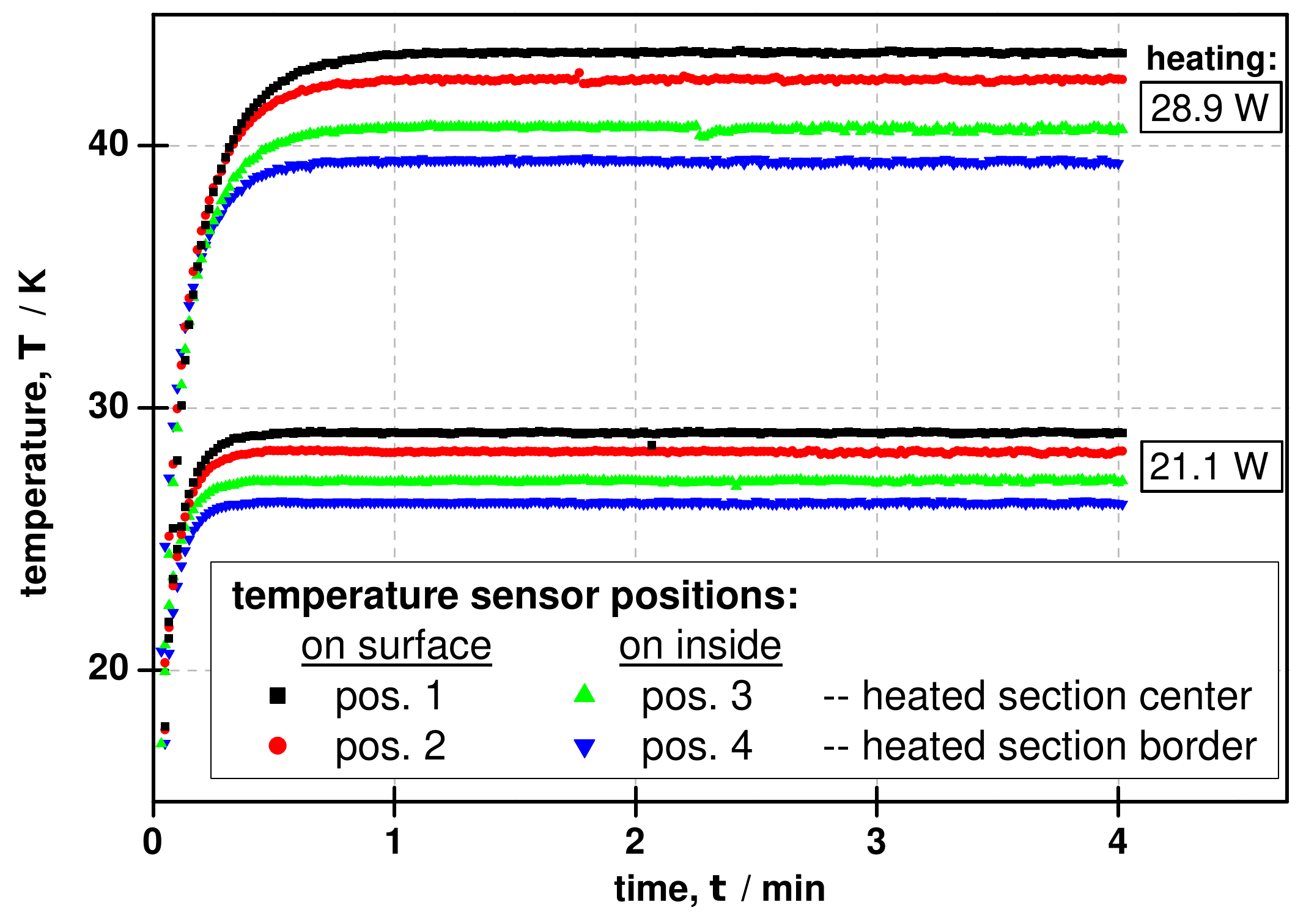}\tabularnewline
\end{tabular}
\par\end{centering}

\caption{Validation of the variable temperature with a CICC dummy consisting
of a copper rod inside a stainless steel tube representing a worst
case scenario of HTS cables. Four Cernox temperature sensors are placed
at different positions. Temperature sensor positions and the layout
of the dummy are shown on in (a: left). The time dependent readout
of the sensors is show for two different heating powers (\SI{21.1}{\watt}
and \SI{28.9}{\watt}) in (b: right).\label{fig:T-variable-insert-dummy}}
\end{figure}

The performance of the variable temperature insert is validated on
a CICC dummy consisting of a copper rod of \SI{8}{\milli\metre}
diameter inside a round stainless steel jacket with \SI{1}{\milli\metre}
of wall thickness. This assembly results in very low radial (transverse)
and high longitudinal thermal conductivities and is therefore the
worst case scenario for the variable temperature insert as heat is
transferred radially from the heating foils to the sample and lost
longitudinally to the helium bath. Four Cernox sensors are used to
measure the temperature of the dummy at different positions. Two sensors
(sensor positions 1 \& 2) are located within the stainless steel jacket,
just below the heating foils, reading the sample's surface temperatures.
The other two temperature sensors (sensor positions 3 \& 4) are embedded
in the dummy's core. The layout of the dummy and the placement of
the sensors is shown schematically in figure~\ref{fig:T-variable-insert-dummy}
(a). In figure~\ref{fig:T-variable-insert-dummy} (b), the readout
of these sensors is shown for different heating powers of the variable
temperature insert.

These temperature sensors of the cable dummy are used to assess the
performance of the variable temperature insert. Response time, stability
and temperature distribution are analyzed.

\subsubsection{Response time}

The response of the variable temperature insert is fast in the investigated
temperature range, due to all materials' low heat capacities. It takes
less than \SI{1}{\minute} to heat up the sample and reach a stable
temperature.

\subsubsection{Stability}

After the heating-up time, the measured temperatures are stable with
flat temperature vs. time curves. Temperature variations at constant
heating power remain below \SI{0.3}{\kelvin} at all investigated
temperatures and can be neglected. Thus, sample temperature are assumed
to be constant after the heat up time.

\subsubsection{Temperature distribution}

A sample is surrounded with heating foils inside the variable temperature
insert. Outside the insert, the sample remains immersed in the liquid
helium bath. Heat is transported along the sample from the heated
section to the helium bath. This heat flux depends on the temperature
difference, the distance, the cross sectional area of the sample and
the thermal conductivities of the constituent materials. Consequently
the sample temperature is not homogenous; there is always a temperature
gradient in the sample. The thermal conductivity of the jacket material
is much lower compared with the conductivity of the central copper
rod. Thus, the thermal conductivity of the cable is significantly
higher in longitudinal direction than in transverse direction. Along
the sample (longitudinal direction) the main part of the heat flux
takes place in the copper center, resulting in a pronounced transverse
temperature gradient. Due to the close proximity to the heating foils,
the two temperature sensors on the surface of the sample (sensor positions
1 \& 2) yield the highest temperatures. Inside the sample, the temperatures
are significantly lower (sensor positions 3 \&4 ). The precedence
of the temperatures (pos. 1 > pos. 2 > pos. 3 > pos. 4) remains true
for the investigated heating powers, just the differences between
the sensors become increasingly pronounced with stronger heating.
The main contribution reducing the temperature is therefore the transverse
distance from the sample's surface followed by the longitudinal distance
from the center of the heated section. The temperature differences
of the two sensors on the sample's surface are less than \SI{\pm 0.7}{\kelvin}
event at the higher heating power. This low differences, justify the
assumption that the sample's surface temperature is homogeneous in
all following investigations. Towards the inside of the sample, the
temperature differences become more pronounced. The sample's inside
sensors register approximately \SI{2}{\kelvin} lower temperatures
than at the corresponding surface positions.

\subsection{Temperature distribution of the Twisted Stacked-Tape Cable\label{sub:T-distribution}}

\begin{figure}[tbh]
\begin{centering}
\includegraphics[width=0.5\textwidth]{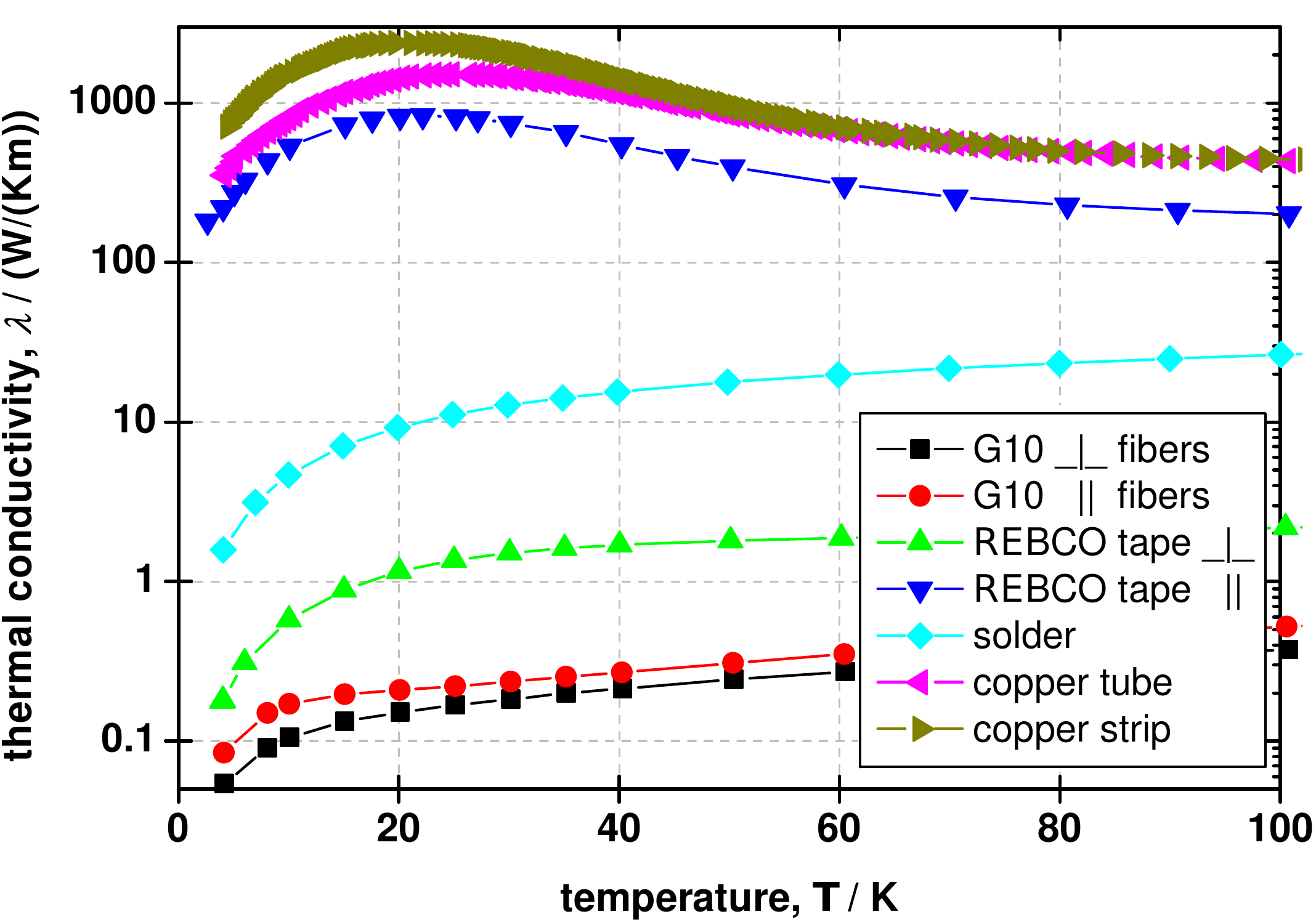}
\par\end{centering}

\caption{Thermal conductivity of the constituent materials of the investigated
Twisted Stacked-Tape Cable sample in the operating temperature range
of the variable temperature insert. Data from \cite{Bagrets2014,Bagrets2014a,Barth-PhD}.\label{fig:TC}}
\end{figure}

There are two Cernox\textsuperscript{\textregistered} sensors are
attached to the sample's surface (one at the upper border of the heated
section at position \SI{-50}{\milli\metre} and the other in the center
of the heated section at position at \SI{0}{\milli\metre}), however,
in contrast to the cable dummy it is not possible to place temperature
sensors inside the Twisted Stacked-Tape Cable sample. This precludes
a fully experimental approach to determine the sample's temperature
distribution. Instead, the temperature distribution is calculated
using Finite-Element-Method (FEM) models with the commercial software
package ``Comsol Multiphysics'' \cite{Comsol-SUST} using the average
of the two surface sensors as an input. In the following, this average
is referred to as the sample's surface temperature \sym{T}{surface}.
Comparing with the HTS cable dummy's experiments (subsection~\ref{sub:T-variable-insert-validation})
it is clear that the temperature variations on the sample's surface
are not pronounced due to the close proximity to the heating foils,
allowing the assumption of a homogenous surface temperature in the
FEM model. The heated section of the sample is modeled at full scale
(see table~\ref{tab:Sample-parameters}) in three dimensions using
the direction and temperature dependent thermal conductivities of
all constituent materials as shown in figure~\ref{fig:TC}. The thermal
conductivity of the \YBCO tapes are mainly influenced in longitudinal
direction by the copper stabilizer and in transverse direction by
the Hastelloy\textsuperscript{\textregistered} substrate resulting
in a three order of magnitude anisotropy \cite[chap. 4.3.5]{Barth-PhD}\@.

\begin{figure}[tb]
\begin{centering}
\includegraphics[width=1\textwidth]{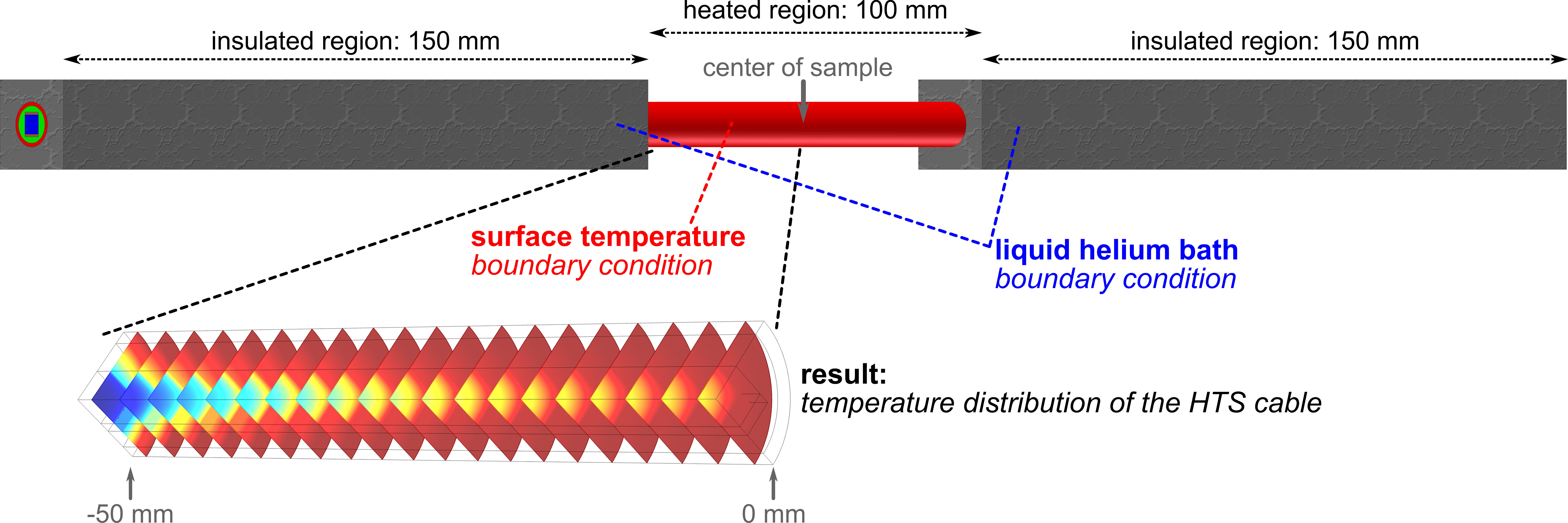}
\par\end{centering}

\caption{Schematic drawing of the FEM models used in the simulations of the
temperature distribution of HTS cables in the variable temperature
insert. An exemplary temperature distribution is shown as a color
gradient from cold (blue) to hot (red). Drawing is not up to scale.\label{fig:T-distribution-method}}
\end{figure}

In the model, temperature sources (the heating foils: length \SI{100}{\milli\metre})
and temperature sinks (the helium bath outside the insulated area:
length: \ensuremath{2\times}\SI{150}{\milli\metre}) are imposed as
boundary conditions. For optimal resolution, different mesh configurations
are used. The \YBCO tapes are netted with a mapped mesh (mesh with
explicit node locations) allowing the placement of several nodes within
the width of each tapes' superconducting layer. A free tetradic mesh,
with much lower node density, is used in the variable temperature
insert. This simulation method is shown schematically in figure~\ref{fig:T-distribution-method}.

\begin{figure}[tb]
\begin{centering}
\includegraphics[width=0.5\textwidth]{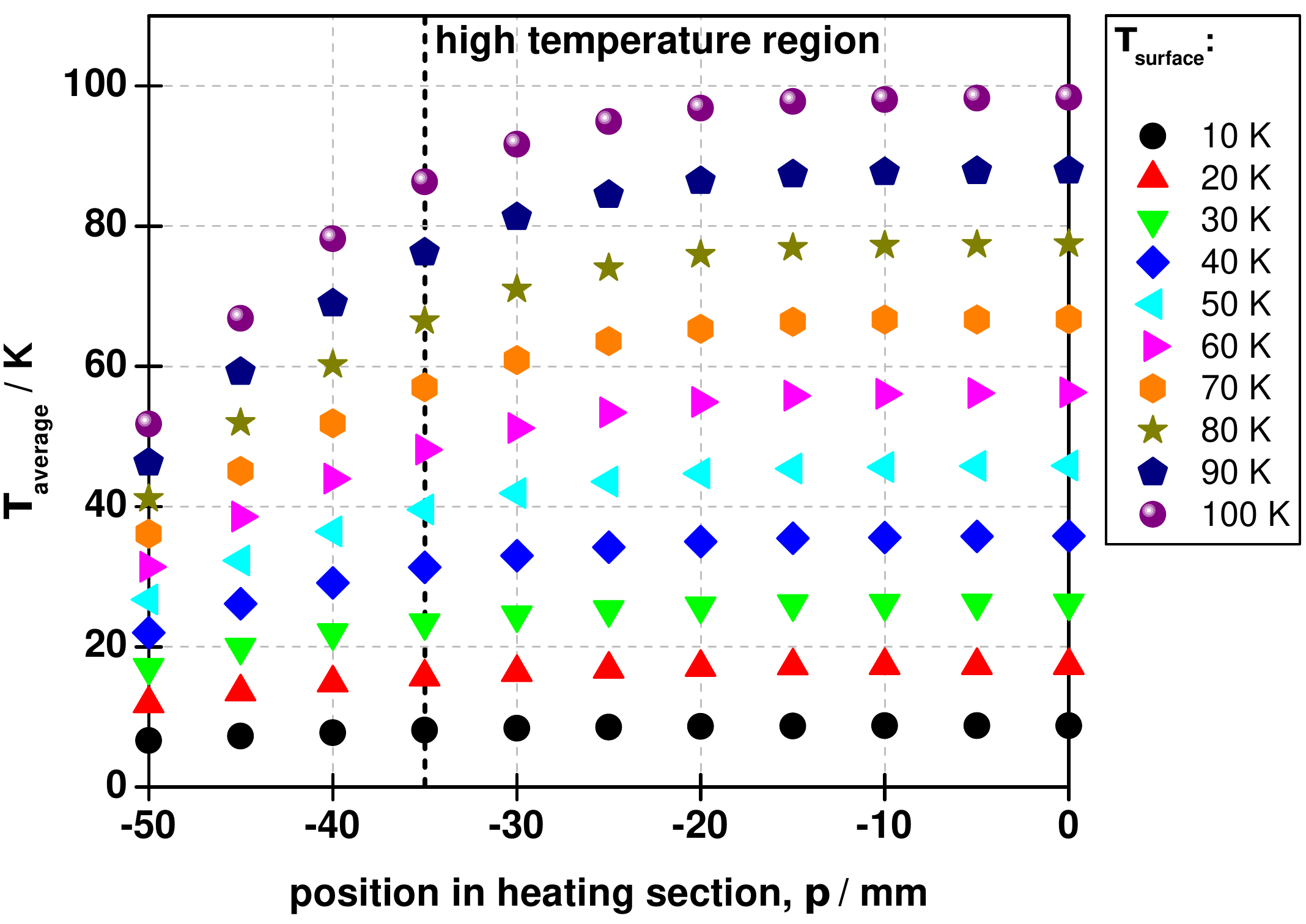}
\par\end{centering}

\caption{Simulated temperature distribution of a $40$-tape TSTC. The average
temperature of the cable \sym{T}{average} is shown at different positions
in the heating section from the border (position: \SI{-50}{\milli\metre})
to the center (position: \SI{0}{\milli\metre}) of the \SI{100}{\milli\metre}
long heating section).\label{fig:T-distribution}}
\end{figure}

The temperature distribution is calculated for surface temperatures
from \SIrange{10}{100}{\kelvin}. The temperatures of the nodes along
the width of all tapes' superconducting layers are averaged each \SI{5}{\milli\metre}
along length of the sample. These temperatures \sym{T}{average} are
shown in figure~\ref{fig:T-distribution} from the border (position:
\SI{-50}{\milli\metre}) to the center of the heating section (position:
\SI{0}{\milli\metre}).

\begin{figure}[!tbph]
\begin{centering}
\includegraphics[width=0.5\textwidth]{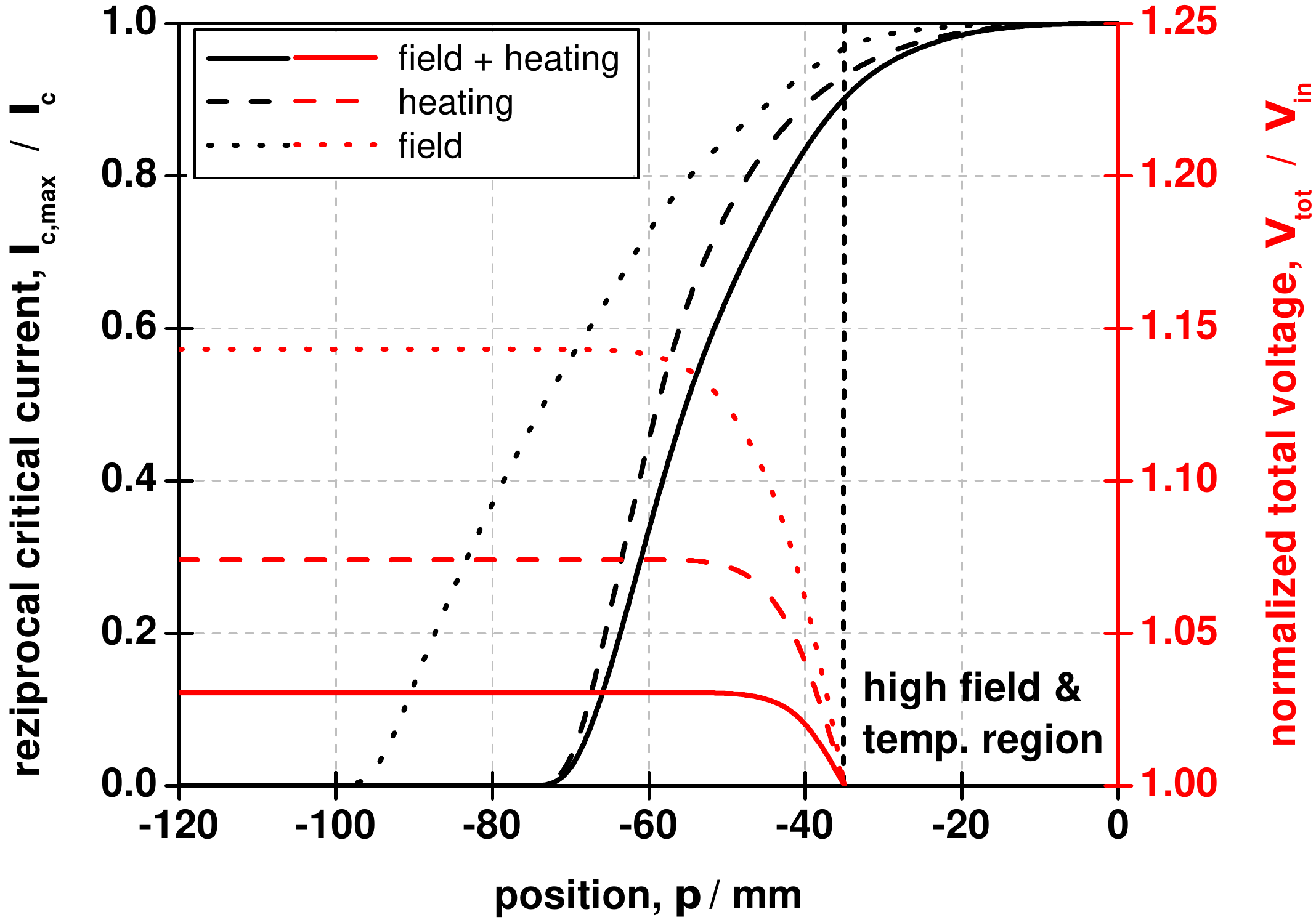}
\par\end{centering}

\caption{Extrapolated reciprocal normalized critical currents (in black) and
normalized total voltages (in red) depending on the position along
the sample for three scenarios: field active (\SI{12}{\tesla} magnetic
background field), heating active (\SI{40}{\kelvin} sample surface
temperature) as well as field and heating active (\SI{12}{\tesla}
magnetic background field at \SI{40}{\kelvin} sample surface temperature).\label{fig:Outside-zone}}

\end{figure}

With the copper sheath and the copper tapes on the top and the bottom
of the superconductor stack, the TSTC contains a lot of highly conductive
material. They result in a significant thermal transport along the
sample and leads to very low average temperatures near the borders
of the heating section. A central \SI{70}{\milli\metre} long region
(from position \SI{-35}{\milli\metre} to \SI{35}{\milli\metre})
with much lower temperature deviation, corresponding to the ``high
field region'' of the test facility's magnet is therefore defined
as the ``high temperature region'' and the active zone of the variable
temperature insert. Within this region, the temperature deviations
are \SI{20.2}{\percent} at \SI{100}{\kelvin} surface temperature
and \SI{<14.4}{\percent} in the surface temperature range of interest
for the experiments (\SIrange{20}{40}{\kelvin}). Outside this region,
the temperature drops quickly. This means that in the temperature
dependent characterization of the TSTC sample (subsection~\ref{sec:T-variable}),
\SI{70}{\milli\metre} well defines the working area of the variable
temperature insert. In this region the temperature varies no more
than \SI{\pm 2.7}{\kelvin}, allowing \YBCO's temperature dependence
of the critical current to be assumed to be linear while averaging
the temperature distribution. This zone, is therefore used to translate
voltages into electric fields regardless of the separation of the
voltage taps. Temperatures and magnetic fields are highest only inside
this zone, limiting the whole sample's current carrying capabilities.
Outside of this zone, temperature and field are decreasing strongly.
The contribution of the parts of the sample outside of the high field
and high temperature region are determined by extrapolating the test
facility's magnetic field profile as well as the simulated temperature
distribution. Using the field- and temperature dependence published
by the tapes' manufacturer, the current carrying capabilities are
calculated depending on the position $p$ along the length of the
sample for three scenarios: field active (\SI{12}{\tesla} magnetic
background field), heating active (\SI{40}{\kelvin} sample surface
temperature) as well as field and heating active (\SI{12}{\tesla}
magnetic background field at \SI{40}{\kelvin} sample surface temperature).
In order to make a worst case assumption, the resulting total voltages
are obtained assuming a power-law behavior with a low n-value of $8$
(lowest n-value observed in all measurements of the TSTC sample, see
figure~\ref{fig:N-values}). In figure~\ref{fig:Outside-zone},
reciprocal normalized critical currents \ensuremath{I_{\n{c,max}}/I_{\n{c}}(p)}
and normalized total voltages \ensuremath{V_{\n{tot}}(p)/V_{\n{in}}}
are shown, the data is symmetric around the sample's center, the zero
position ($p=0$). 

The parts of the sample contribute \SI{14}{\percent} if only the
magnet is used, \SI{7}{\percent} if solely the variable temperature
insert is activated and \SI{3}{\percent} in the case of magnetic
background field and heating. These low contributions, especially
in the case of magnetic background fields and heating, justify to
neglect the parts of the sample outside the high field and high temperature
zone. In these cases, voltages are only generated in the sample's
central \SI{70}{\milli\metre} and this distance is used to convert
the obtained voltages into electric fields.

\section{Results}

\subsection{Critical current of magnetic field cycles\label{sec:Load-cycling}}

Initially, the cable is checked for degradation of the current carrying
capabilities due to high transverse mechanical loads. This is achieved
through cycling of the magnetic background field, currents and transverse
loads. The magnetic background field is increased from zero field
to \SI{12}{\tesla}. Every \SI{2}{\tesla}, the critical current is
measured at \SI{4.2}{\kelvin}.

\begin{figure}[tbh]
\begin{centering}
\includegraphics[width=0.5\textwidth]{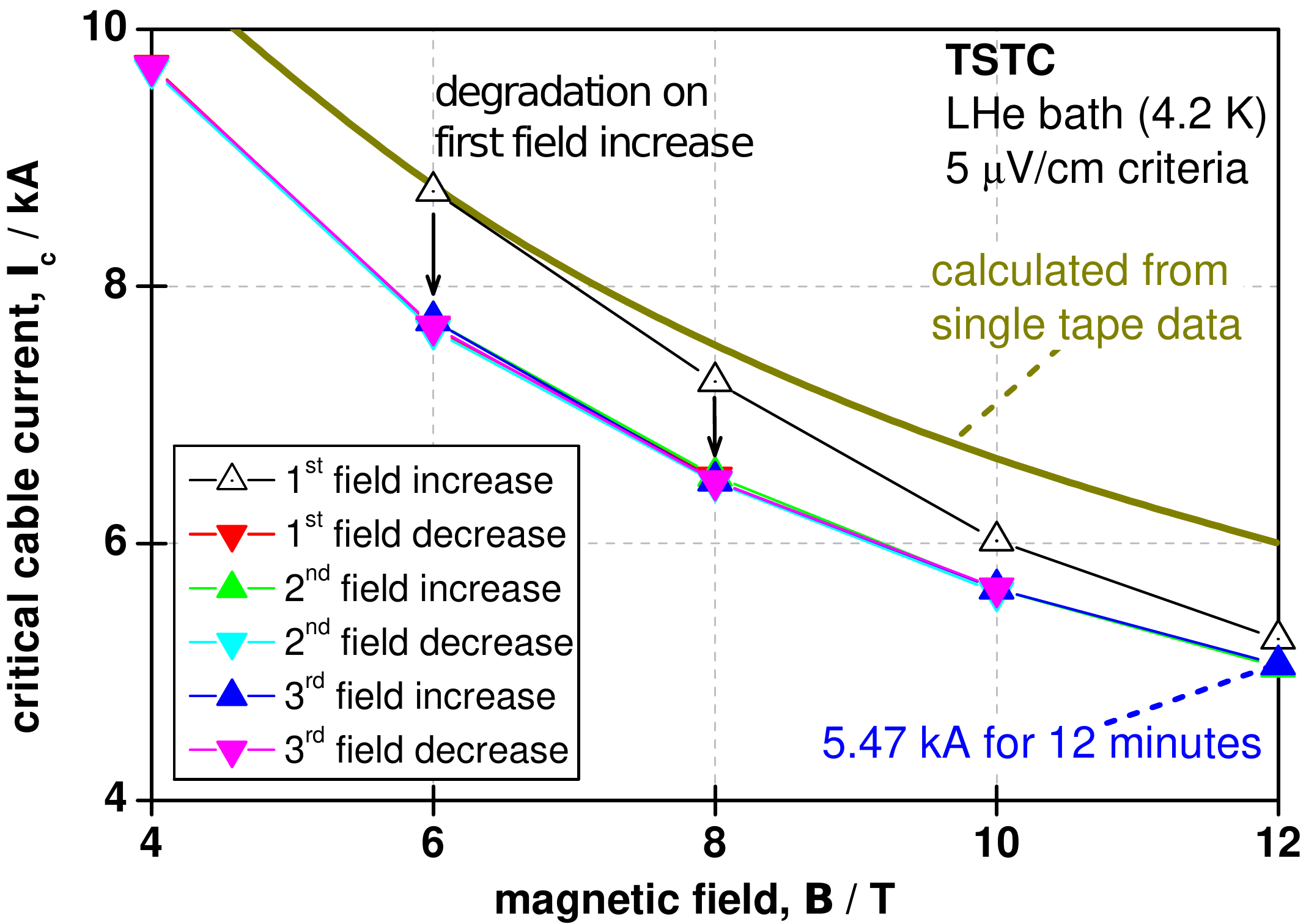}
\par\end{centering}

\caption{Critical current measurements at increasing and decreasing magnetic
background fields of a $40$-tape TSTC. The voltage tap pair with
maximal separation is used; linear ohmic contribution in the V-I curves
is subtracted. Field dependence of the critical currents (yellow curve)
is calculated based on single tape data from \cite{Hazelton2012-Napa-Workshop}
and normalized at \SI{6}{\tesla}.\label{fig:Load-cycling}}
\end{figure}

In the \SIrange{8}{12}{\tesla} range, the current carrying capabilities
of SuperPower advanced pinning \YBCO tapes are only slightly reduced
with magnetic field increases. Consequently, the Lorentz forces, and
the transverse loads on the cable, are maximal at \SI{12}{\tesla}.
After reaching maximal field, and maximal transverse loads, the background
field is reduced back to zero. Once again, the critical currents are
measured every \SI{2}{\tesla}. Any occurring degradation is exposed
by comparing the critical currents measured at identical background
fields on field increases and decreases. This procedure is repeated
several times to detect further degradation of the sample. The measured
critical currents are shown in figure~\ref{fig:Load-cycling}. In
these measurements, the sample is kept at \SI{4.2}{\kelvin}, the
variable temperature insert is not activated. The voltage tap pair
on the copper jacket with maximal separation (see figure~\ref{fig:FBI-t-variable-insert})
is used to average over the behavior of all the tapes. The sample
includes two clamped BSCCO - \YBCO connectors which are a very compact
way of connecting the 40 \YBCO tapes. However, as the contact resistances
are not perfectly homogeneous, the sample's tapes carry different
fractions of the total current. These differences decrease with increasing
voltage during the superconducting transition. A \SI{5}{\micro\volt\per\centi\metre}
criteria is therefore used to determine the critical currents in all
in-field measurements of the TSTC sample. Furthermore, the inhomogeneous
contact resistances and the following current redistribution result
in a linear ohmic contribution during the critical current measurements.
These are subtracted from the V-I curves before determining the critical
currents. In this figure a field dependence of the critical currents
calculated based on single tape data from \cite{Hazelton2012-Napa-Workshop}
is shown between \SI{6}{\tesla} and \SI{12}{\tesla} by normalizing
at \SI{6}{\tesla}. 

On the first increase of the magnetic background field, the current
carrying capabilities of the sample degrade. At \SI{6}{\tesla}, the
sample current is exactly as to be expected from the single tape tape.
To higher magnetic background fields, the Twisted Stacked-Tape Cable
sample carries lower current relative to the single tape data. The
high Lorentz forces, step-by-step degrade the superconductor tapes
with increasing fields. These degradations however do not continue
during following magnetic field cycles. After the first cycle, the
current carrying capabilities remain unchanged even if the TSTC is
kept at the maximum current of \SI{5.47}{\kilo\ampere} at \SI{12}{\tesla}
for \SI{12}{\minute}. The critical currents measured on the first
increase of the magnetic field and on later field cycles are shown
in table~\ref{tab:TSTC-degradation}. The current carrying capabilities
of all magnetic field cycles but the first field increase are averaged.

\begin{table}[tbh]
\caption{Degradation of a $40$-tape Twisted Stacked-Tape Cable due to transverse
mechanical loads at high magnetic background fields. Critical currents
of field cycles, following on the first magnetic field increase, are
averaged.\label{tab:TSTC-degradation}}

\centering{}%
\begin{tabular}{cccc}
\toprule 
{\footnotesize{}field} & {\footnotesize{}critical current - first increase} & {\footnotesize{}critical current - later} & {\footnotesize{}degradation}\tabularnewline
\midrule
\midrule 
{\footnotesize{}{\footnotesize{}\SI{6}{\tesla}}} & {\footnotesize{}{\footnotesize{}\SI{8.74}{\kilo\ampere}}} & {\footnotesize{}{\footnotesize{}\SI{7.70}{\kilo\ampere}}} & {\footnotesize{}{\footnotesize{}\SI{11.9}{\percent}}}\tabularnewline
\midrule 
{\footnotesize{}{\footnotesize{}\SI{8}{\tesla}}} & {\footnotesize{}{\footnotesize{}\SI{7.26}{\kilo\ampere}}} & {\footnotesize{}{\footnotesize{}\SI{6.50}{\kilo\ampere}}} & {\footnotesize{}{\footnotesize{}\SI{10.5}{\percent}}}\tabularnewline
\midrule 
{\footnotesize{}{\footnotesize{}\SI{10}{\tesla}}} & {\footnotesize{}{\footnotesize{}\SI{6.02}{\kilo\ampere}}} & {\footnotesize{}{\footnotesize{}\SI{5.64}{\kilo\ampere}}} & {\footnotesize{}{\footnotesize{}\SI{6.3}{\percent}}}\tabularnewline
\midrule 
{\footnotesize{}{\footnotesize{}\SI{12}{\tesla}}} & {\footnotesize{}{\footnotesize{}\SI{5.26}{\kilo\ampere}}} & {\footnotesize{}{\footnotesize{}\SI{5.05}{\kilo\ampere}}} & {\footnotesize{}{\footnotesize{}\SI{4.0}{\percent}}}\tabularnewline
\bottomrule
\end{tabular}
\end{table}

Determination of degradation is not possible at fields of \SI{4}{\tesla}
and below as the critical currents of the sample are above the maximal
current of the test facility (\SI{10}{\kilo\ampere}). At background
fields of \SI{6}{\tesla}, a maximal degradation of \SI{11.9}{\percent}
is obtained. For higher fields, the degradation decreases. The non-constant
degradation implies that damage to the sample occurs not just during
the critical current measurement at maximal field. Some degradation
already occurs below maximal magnetic fields. The factor of degradation
is close at \SI{6}{\tesla} and \SI{8}{\tesla}, signifying that the
biggest fraction of the total degradation is occurring at higher magnetic
background fields.

Transverse mechanical loads can degrade the \YBCO tapes in TSTCs,
these degradations are described above. In the investigated sample,
the hollow space between the stack of superconductor tapes and the
copper jacket is quite well filled with soft solder by liquifying
and injecting the solder through holes in the jacket while the sample
is heated on a heating plate. However, the solder did not support
well the Lorentz forces, allowing a movement of the stack leading
to a degradation of the superconductor tapes. Due to the superior
mechanical properties and to possibility to be drawn into the sample
with vacuum would make low thermal expansion epoxy resin mixtures
superior filling agents which may have prevented such sample degradations
\cite{Barth2013-epoxy}.

\subsection{Temperature dependence at different background fields\label{sec:T-variable}}

\begin{figure}[tbh]
\begin{centering}
\includegraphics[width=0.5\textwidth]{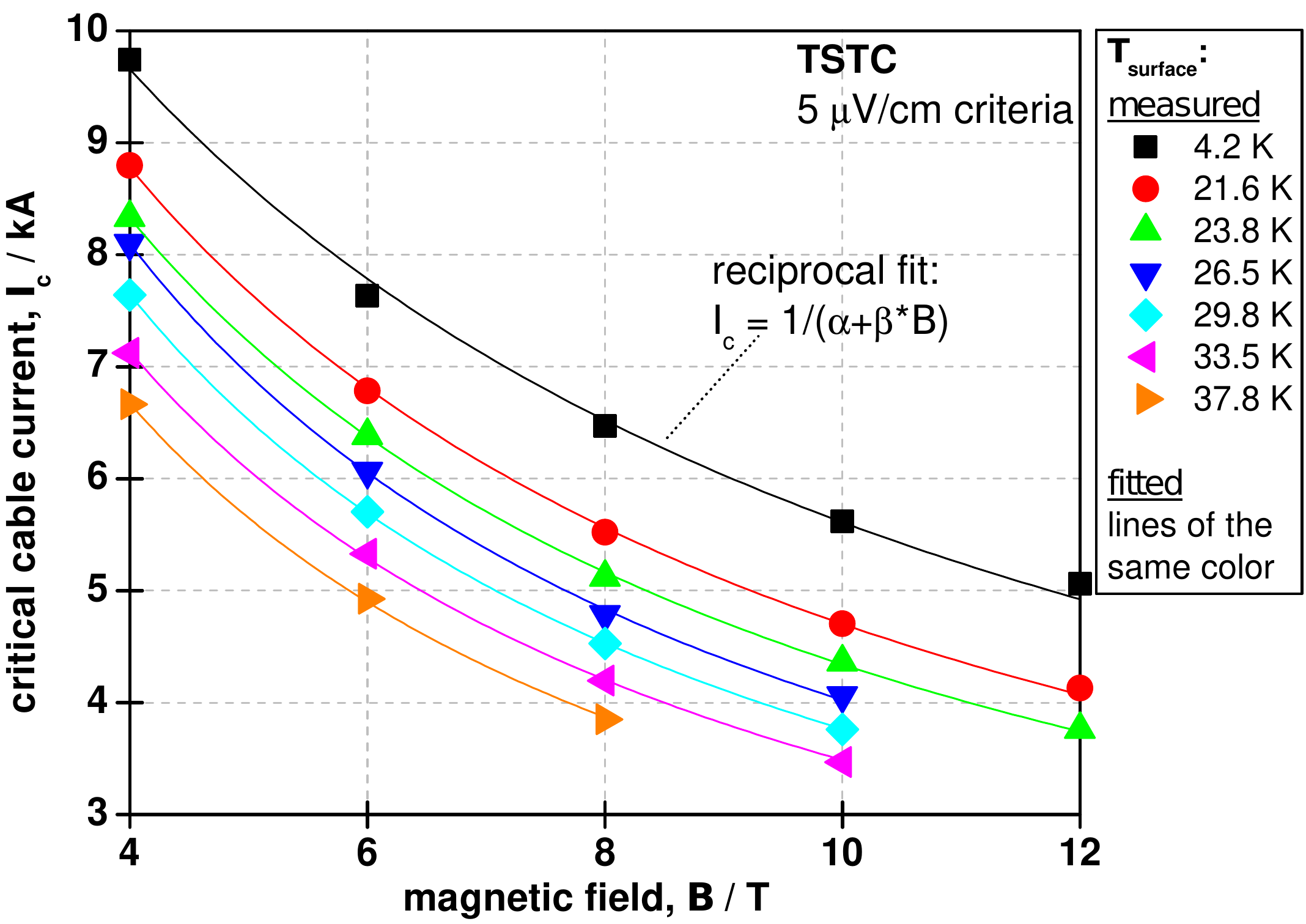}
\par\end{centering}

\caption{Field and surface temperature dependent measurements of a 40-tape
TSTC. The measurements (points) use the voltage taps on the copper
jacket with maximal separation. Linear ohmic contribution in the V-I
curves is subtracted. Critical currents are determined with the \SI{5}{\micro\volt\per\centi\metre}
criteria assuming a active region of \SI{70}{\milli\metre} for field
and temperatures. Data is fitted with reciprocal functions (lines).\label{fig:T-variable}}
\end{figure}

In this part of the TSTC sample characterization, the variable temperature
insert of the FBI test facility is utilized. The critical currents
are measured in each of the \SI{2}{\tesla} magnetic background field
steps for different cable surface temperatures. It was attempted to
obtain the same sample surface temperature steps in the whole magnetic
field range. The two copper tapes on the top and the bottom of the
\YBCO tapes stack, the solder and the encasing copper tube electrically
stabilize the sample. The stabilization is sufficient, no thermal
runaway is observed, even during superconducting transitions close
to\SI{10}{\kilo\ampere} transport current. Within the heated section,
two temperature sensors are attached to the sample's surface (one
at the upper border of the heated section at position \SI{-50}{\milli\metre}
and the other in the center of the heated section at position at \SI{0}{\milli\metre}).
As expected from the validation of the variable temperature insert
(subsection~\ref{sub:T-variable-insert-validation}), their readouts
are very similar (within \SI{\pm 0.9}{\kelvin} in the investigated
range of heating powers) and their average value is therefore referred
to as the samples surface temperature \sym{T}{surface}. In figure~\ref{fig:T-variable},
the current carrying capabilities of the TSTC sample are shown for
sample surface temperatures from \SI{4.2}{\kelvin} to \SI{37.8}{\kelvin}
using a \SI{5}{\micro\volt\per\centi\metre} criteria assuming an
active region of the magnetic field and the temperature of \SI{70}{\milli\metre}.
Linear ohmic contribution in the V-I curves is subtracted. From the
experimental point of view it has to be noted that the twist pitch
of the sample is with \SI{200}{\milli\metre} (see table~\ref{tab:Sample-parameters})
larger than the \SI{70}{\milli\metre} long high field and high temperature
region. The tapes are therefore not exposed to all orientations of
the magnetic background field adding some uncertainty to the gained
field dependence due to \emph{RE}BCO's anisotropy.

\begin{figure}[tbh]
\begin{centering}
\includegraphics[width=0.5\textwidth]{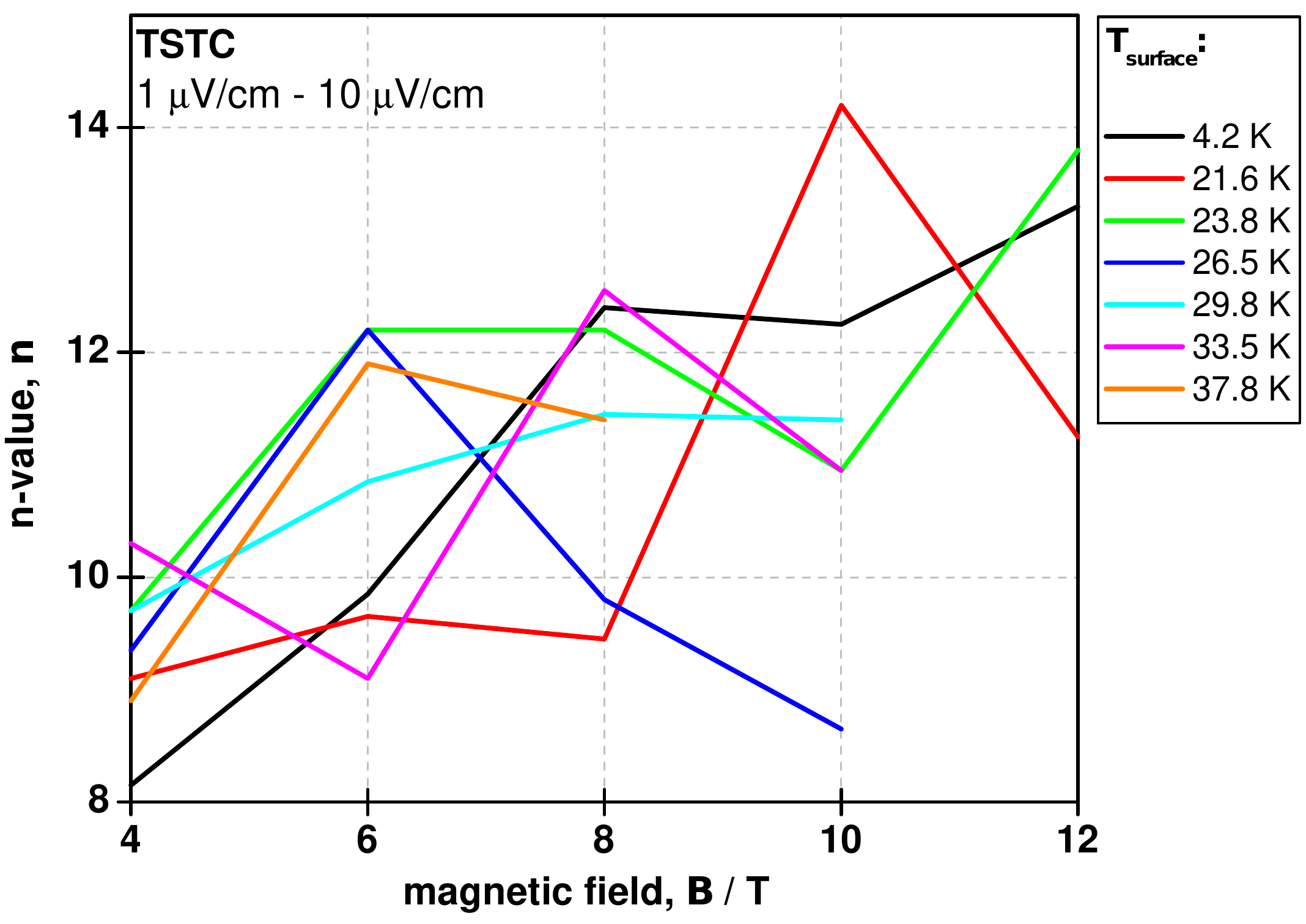}
\par\end{centering}

\caption{The n-values of the superconducting transitions of a $40$ tape TSTC.
The measurements are of the degraded cable at different cable temperatures
and magnetic background fields. The n-values are calculated in the
electric field range \SIrange{1}{10}{\micro\volt\per\centi\metre}
using the voltage tap pair on the copper jacket with maximal separation.
Linear ohmic contribution in the V-I curves is subtracted.\label{fig:N-values}}
\end{figure}

The curves of different surface temperatures are regularly spaced
and are of reciprocal shape (\ensuremath{\sym{I}{c} = \frac{1}{\left(\alpha+\beta\cdot B\right)}}).
The n-values are calculated for all superconducting transitions of
the field and temperature dependent measurements. The calculations
are done in the electric field range from \SI{1}{\micro\volt\per\centi\metre}
to \SI{10}{\micro\volt\per\centi\metre}. The same voltage taps are
used, assuming an active region of field and temperature of \SI{70}{\milli\metre}.
The n-values are shown in figure~\ref{fig:N-values}.

As could be expected from the sample layout, all n-values are low,
they are in the $8$ to $14$ range. There are two aspects contributing
to the low n-values. Firstly, the inhomogeneous contact resistances
of the two \YBCO - BSCCO connectors, and the resulting current redistribution.
The tapes do not carry the same fraction of the total current and
therefore begin their transition from superconducting state to normal
conduction at different cable currents. As the voltage is read at
the copper sheath it corresponds to an averaging of the behavior of
all tapes. This ``smears'' the transition and lowers the n-values.Secondly,
the degradation of the sample, which occurred on the first field increase.
Some of the \YBCO tapes were damaged, reducing the current they can
carry. This damage strongly in-homogenizes the sample, causing the
tapes to generate voltages at very low cable currents. In a non-degraded
TSTC, significantly higher n-values are therefore to be expected.

Using the temperature distribution obtained in the FEM model of the
TSTC sample (subsection~\ref{sub:T-distribution}), average sample
temperatures \sym{T}{avg} are calculated for the measured surface
temperatures \sym{T}{surface} by averaging the distribution in the
\SI{70}{\milli\metre} long high temperature region. These temperatures
as well as the statistic (from the averaging and the FEM model) and
the systematic (from the temperature measurement) temperature uncertainty
are shown in table~\ref{tab:Tavg-Tsurface}. 

\begin{table}[!tbph]
\centering{}\caption{Sample surface temperatures \sym{T}{surface}, average sample temperatures
\sym{T}{avg}, statistic and systematic measurement uncertainty derived
from the FEM model of the TSTC sample.\label{tab:Tavg-Tsurface}}
{\scriptsize{}}%
\begin{tabular}{cccc}
\toprule 
{\scriptsize{}surface temperature, \sym{T}{surface}} & {\scriptsize{}average temperature, \sym{T}{average}} & {\scriptsize{}statistic uncertainty} & {\scriptsize{}systematic uncertainty}\tabularnewline
\midrule
\midrule 
{\scriptsize{}{\scriptsize{}\SI{4.2}{\kelvin}}} & {\scriptsize{}{\scriptsize{}\SI{4.2}{\kelvin}}} & {\scriptsize{}-} & {\scriptsize{}-}\tabularnewline
\midrule 
{\scriptsize{}{\scriptsize{}\SI{21.6}{\kelvin}}} & {\scriptsize{}{\scriptsize{}\SI{18.3}{\kelvin}}} & {\scriptsize{}{\scriptsize{}\SI{\pm 0.6}{\kelvin}}} & {\scriptsize{}{\scriptsize{}\SI{\pm 0.3}{\kelvin}}}\tabularnewline
\midrule 
{\scriptsize{}{\scriptsize{}\SI{23.8}{\kelvin}}} & {\scriptsize{}{\scriptsize{}\SI{20.2}{\kelvin}}} & {\scriptsize{}{\scriptsize{}\SI{\pm 0.9}{\kelvin}}} & {\scriptsize{}{\scriptsize{}\SI{\pm 0.4}{\kelvin}}}\tabularnewline
\midrule 
{\scriptsize{}{\scriptsize{}\SI{26.5}{\kelvin}}} & {\scriptsize{}{\scriptsize{}\SI{22.5}{\kelvin}}} & {\scriptsize{}{\scriptsize{}\SI{\pm 1.0}{\kelvin}}} & {\scriptsize{}{\scriptsize{}\SI{\pm 0.4}{\kelvin}}}\tabularnewline
\midrule 
{\scriptsize{}{\scriptsize{}\SI{29.8}{\kelvin}}} & {\scriptsize{}{\scriptsize{}\SI{25.4}{\kelvin}}} & {\scriptsize{}{\scriptsize{}\SI{\pm 1.1}{\kelvin}}} & {\scriptsize{}{\scriptsize{}\SI{\pm 0.6}{\kelvin}}}\tabularnewline
\midrule 
{\scriptsize{}{\scriptsize{}\SI{33.5}{\kelvin}}} & {\scriptsize{}{\scriptsize{}\SI{28.9}{\kelvin}}} & {\scriptsize{}{\scriptsize{}\SI{\pm 1.3}{\kelvin}}} & {\scriptsize{}{\scriptsize{}\SI{\pm 0.7}{\kelvin}}}\tabularnewline
\midrule 
{\scriptsize{}{\scriptsize{}\SI{37.8}{\kelvin}}} & {\scriptsize{}{\scriptsize{}\SI{32.7}{\kelvin}}} & {\scriptsize{}{\scriptsize{}\SI{\pm 1.6}{\kelvin}}} & {\scriptsize{}{\scriptsize{}\SI{\pm 0.9}{\kelvin}}}\tabularnewline
\bottomrule
\end{tabular}
\end{table}

\subsection{Characterization at \SI{77}{\kelvin}, self-field after the in-field
measurements\label{sec:77K-characterization}}

Following the described magnetic field and temperature tests at the
FBI facility, the tested sample was sent back to MIT, and the critical
currents were tested at \SI{77}{\kelvin} in a liquid nitrogen bath
without a magnetic background field (self-field conditions) \cite{Chiesa2014}.
Two regions of the conductor were tested: one at the center section
which was tested in the high field at the FBI and the other near the
end of the conductor which was not exposed to the high field. The
critical current test results were \SI{1.70}{\kilo\ampere} at the
\SI{1}{\micro\volt\per\centi\metre} criterion with an n-value of
$12$ for the center region and \SI{1.99}{\kilo\ampere} with an n-value
of $26$ at the end region. It indicates the center region had degraded
by \SI{15}{\percent} during the high field test at the FBI test facility,
the n-value of the center region is in very good agreement with with
the n-value range observed during the field- and temperature dependent
measurements. It is also noted that the linear ohmic contribution
in the V-I curve was as small as \SI{2.1e-9}{\ohm}, less than that
observed at the high field experiments at the FBI test. A larger linear
ohmic component during the FBI tests is to be expected as the active
region of fields and temperatures is with \SI{70}{\milli\metre} rather
small. Thus, the voltage during the superconducting transitions are
low, making the current redistribution due to the contact resistance
inhomogeneities more pronounced. Transverse load test performed after
the critical current tests at \SI{77}{\kelvin} was shown additional
degradation by only \SI{3}{\percent} for the center region, on the
other hand by \SI{20}{\percent} for the end region by the external
transverse load of \SI{400}{\kilo\newton\per\metre}. Further transverse
load test results can be obtained in \cite{Chiesa2014}.

\section{Discussion and conclusion\label{sec:Conclusion}}

The performance of the variable temperature insert of the FBI test
facility has been evaluated with a CICC dummy. Temperature sensors
on the surface and the inside of the dummy are used to determine,
the response time, the stability and the temperature distribution
of the dummy. In the temperature range of interest in less than one
minute, the temperatures are stable, however there is a strong temperature
gradient from the surface to the inside of the sample. The temperature
readouts of the sensors on the surface are similar, allowing the use
their average in the following. As it is not possible to place sensors
on the inside of the TSTC sample, it is modeled in 3D in full scale
using the temperature and direction dependent properties of its constituent
materials. From this FEM model the working area of the variable temperature
insert is determined to be \SI{70}{\milli\metre} with a temperature
variation of less than \SI{14.4}{\percent}. In the utilized temperature
range this corresponds to deviations of less than \SI{\pm 2.7}{\kelvin}
from the average temperature. Averaging assumes a linear temperature
dependence of the critical current for the \YBCO tapes, which is
acceptable in such small temperature range, far below the critical
temperature. Outside this high temperature zone the sample temperature
drops quickly. The zone corresponds to the high field region of the
magnet, therefore this \SI{70}{\milli\metre} are used to transform
voltages into electric fields in all FBI measurements of the TSTC
sample regardless of the separation of the voltage taps. The current
carrying capabilities of the the 40-tape Twisted Stacked Tape Cable
sample are tested at various magnetic background fields of up to \SI{12}{\tesla}.
At higher magnetic fields, the enormous Lorentz forces of up to \SI{65.6}{\kilo\newton\per\metre}
result in an irreversible degradation of the sample's transport current.
In the following field cycles, the degradation does not continue,
it saturates at \SI{11.8}{\percent}. Using the variable temperature
insert, the current carrying capabilities of the sample are measured
in up to \SI{12}{\tesla} magnetic background fields and sample surface
temperatures of up to \SI{37.8}{\kelvin}. Using the temperature distribution
obtained in the FEM model, the average sample temperature and the
temperature uncertainty is calculated. Critical current index n-values
in the range of $8-14$ are obtained during these measurements. Two
separate aspects are likely contributing to the low n-values: Firstly,
and secondly the inhomogeneous contact resistances of the sample's
clamped \YBCO - BSCCO connectors which cause the tapes to carry different
fractions of the total current and to start their superconducting
transitions at different cable currents.. Secondly, the sample's degradation,
meaning damage to some of the \YBCO tapes resulting in strong current
redistribution already at low cable currents. Higher n-values are
to be expected in non-degraded TSTCs. As the active zone of the fields
and temperatures (\SI{70}{\milli\metre}) is lower the TSTC sample's
twist pitch (\SI{200}{\milli\metre}), we cannot compare the cable's
performance with the field- and temperature dependence of single tapes'
critical currents. Following on the magnetic field and temperature
tests at the FBI facility, the tested sample is characterized at \SI{77}{\kelvin},
self-field. The center region of the cable, which was exposed to the
high transverse Lorentz forces during the in-field measurements, the
current carrying capabilities are reduced by \SI{15}{\percent}. This
corresponds well to the behavior observed during the magnetic field
cycles.

\ack{}{}

The authors acknowledge Frank Gr\"{o}ner, Valentin Tschan,
Sascha Westenfelder and Anton Lingor. Only their technical
support made these experiments possible.

\bibliographystyle{unsrt}
\addcontentsline{toc}{section}{\refname}\bibliography{library}

\begin{thebibliography}{10}

\bibitem{Goldacker2006}
W~Goldacker, R~Nast, G~Kotzyba, S~I Schlachter, a~Frank, B~Ringsdorf,
  C~Schmidt, and P~Komarek.
\newblock {High current DyBCO-ROEBEL Assembled Coated Conductor (RACC)}.
\newblock {\em Journal of Physics: Conference Series}, 43:901--904, June 2006.

\bibitem{Long2010}
N~J Long, R~A Badcock, K~Hamilton, A~Wright, Z~Jiang, and L~S Lakshmi.
\newblock {Development of YBCO Roebel cables for high current transport and low
  AC loss applications}.
\newblock {\em Journal of Physics: Conference Series}, 234:022021, 2010.

\bibitem{Long2011}
N~J Long.
\newblock {HTS Roebel cables}.
\newblock In {\em HTS4Fusion Conductor Workshop}, KIT CN, Karlsruhe, 2011.

\bibitem{Long2012}
N~J Long, R~A Badcock, C~W Bumby, Z~Jiang, and R~G Buckley.
\newblock {Production and characterisation of HTS roebel cable}.
\newblock In M~Miryala, editor, {\em Superconductivity: Recent Developments and
  New Production Technologies}, chapter~13, pages 259--288. Nova Science
  Publishers, Inc., Huntington, 2012.

\bibitem{Schlachter2011}
S~I Schlachter, W~Goldacker, F~Grilli, R~Heller, and A~Kudymow.
\newblock {Coated Conductor Rutherford Cables (CCRC) for High-Current
  Applications: Concept and Properties}.
\newblock {\em IEEE Transactions on Applied Superconductivity},
  21(3):3021--3024, June 2011.

\bibitem{Kario2013}
A~Kario, M~Vojenciak, F~Grilli, A~Kling, B~Ringsdorf, U~Walschburger, S~I
  Schlachter, and W~Goldacker.
\newblock {Investigation of a Rutherford cable using coated conductor Roebel
  cables as strands}.
\newblock {\em Superconductor Science and Technology}, 26(8):085019, August
  2013.

\bibitem{VanderLaan2009}
D~C van~der Laan.
\newblock {YBa2Cu3O7-delta coated conductor cabling for low ac-loss and
  high-field magnet applications}.
\newblock {\em Superconductor Science and Technology}, 22(6):065013, June 2009.

\bibitem{VanderLaan2011a}
D~C van~der Laan, X~F Lu, and L~F Goodrich.
\newblock {Compact
  GdBa\textsubscript{2}Cu\textsubscript{3}O\textsubscript{7-\ensuremath{\delta}}\
  coated conductor cables for electric power transmission and magnet
  applications}.
\newblock {\em Superconductor Science and Technology}, 24(4):042001, April
  2011.

\bibitem{VanderLaan2012a}
D~C van~der Laan, L~F Goodrich, and T~J Haugan.
\newblock {High-current dc power transmission in flexible
  RE-Ba\textsubscript{2}Cu\textsubscript{3}O\textsubscript{7-\ensuremath{delta}}
  coated conductor cables}.
\newblock {\em Superconductor Science and Technology}, 25(1):014003, January
  2012.

\bibitem{VanderLaan2013}
D~C van~der Laan, P~D Noyes, G~E Miller, H~W Weijers, and G~P Willering.
\newblock {Characterization of a high-temperature superconducting conductor on
  round core cables in magnetic fields up to 20 T}.
\newblock {\em Superconductor Science and Technology}, 26(4):045005, April
  2013.

\bibitem{Takayasu2009}
M~Takayasu, J~V Minervini, and L~Bromberg.
\newblock {US patent: Superconductor Cable (patent number: 8 437 819 B2)},
  2009.

\bibitem{takayasu2011}
M~Takayasu, L~Chiesa, L~Bromberg, and J~V Minervini.
\newblock {Cabling Method for High Current Conductors Made of HTS Tapes}.
\newblock {\em Applied Superconductivity, IEEE Transactions on},
  21(3):2340--2344, 2011.

\bibitem{Takayasu2012}
M~Takayasu, J~V Minervini, L~Bromberg, M~K Rudziak, and T~Wong.
\newblock {Investigation of Twisted Stacked-Tape Cable Conductor}.
\newblock {\em AIP Conference Proceedings}, 1435:273--280, 2012.

\bibitem{Takayasu2012a}
M~Takayasu, L~Chiesa, L~Bromberg, and J~V Minervini.
\newblock {HTS twisted stacked-tape cable conductor}.
\newblock {\em Superconductor Science and Technology}, 25(1):014011, January
  2012.

\bibitem{Chiesa2014}
L~Chiesa, N~C Allen, and M~Takayasu.
\newblock {Electromechanical Investigation of 2G HTS Twisted Stacked-Tape Cable
  Conductors}.
\newblock {\em IEEE Transactions on Applied Superconductivity}, 24(3):6600405,
  June 2014.

\bibitem{Celentano2014}
G~Celentano, G~De Marzi, F~Fabbri, L~Muzzi, G~Tomassetti, A~Anemona,
  S~Chiarelli, M~Seri, A~Bragagni, and A~della Corte.
\newblock {Design of an Industrially Feasible Twisted-Stack HTS
  Cable-in-Conduit Conductor for Fusion Application}.
\newblock {\em IEEE Transactions on Applied Superconductivity}, 24(3):4601805,
  2014.

\bibitem{Uglietti2013}
D~Uglietti, R~Wesche, and P~Bruzzone.
\newblock {Fabrication Trials of Round Strands Composed of Coated Conductor
  Tapes}.
\newblock {\em IEEE Transactions on Applied Superconductivity}, 23(3):4802104,
  2013.

\bibitem{Barth-PhD}
C~Barth.
\newblock {\em {High Temperature Superconductor Cable Concepts for Fusion
  Magnets}}.
\newblock KIT Scientific Publishing, Karlsruhe, 1. edition, 2013.

\bibitem{Bayer2014}
C~M Bayer, C~Barth, P~V Gade, K~Weiss, and R~Heller.
\newblock {FBI Measurement Facility for High Temperature Superconducting Cable
  Designs}.
\newblock {\em IEEE Transactions on Applied Superconductivity}, 24(3):9500604,
  2014.

\bibitem{Barth2011-MEM}
C~Barth, K-P Weiss, W~Goldacker, and S~I Schlachter.
\newblock {Electro-magnetic measurements of coated conductor cables at
  different temperatures}.
\newblock In {\em MEM11 Workshop}, Okinawa, 2011.

\bibitem{Cernox-SUST}
{Lakeshore\textsuperscript{\textregistered} Cryotronics Inc.}
\newblock {Cernox\textsuperscript{\texttrademark} temperature sensor}.

\bibitem{Comsol-SUST}
{COMSOL Inc.}
\newblock {COMSOL Multiphysicstextsuperscript{\textregistered}}.

\bibitem{Bagrets2014}
N~Bagrets, C~Barth, and K~Weiss.
\newblock {Low Temperature Thermal and Thermo-Mechanical Properties of Soft
  Solders for Superconducting Applications}.
\newblock {\em IEEE Transactions on Applied Superconductivity}, 24(3):7800203,
  2014.

\bibitem{Bagrets2014a}
N~Bagrets, W~Goldacker, S~I Schlachter, and C~Barth.
\newblock {Thermal properties of 2G coated conductor cable materials}.
\newblock {\em CRYOGENICS}, 61:8--14, 2014.

\bibitem{Hazelton2012-Napa-Workshop}
D~W Hazelton.
\newblock {2G HTS Conductors at SuperPower}.
\newblock In {\em Low Temperature High Field Superconductor Workshop}, Napa,
  2012. SuperPower.

\bibitem{Barth2013-epoxy}
C~Barth, N~Bagrets, K-P Weiss, C~M Bayer, and T~Bast.
\newblock {Degradation free epoxy impregnation of REBCO coils and cables}.
\newblock {\em Superconductor Science and Technology}, 26(5):055007, 2013.

\end{thebibliography}

\end{document}